\documentclass{article}

\usepackage[eandd, preprint]{neurips_2026}

\usepackage[utf8]{inputenc}
\usepackage[T1]{fontenc}
\usepackage[hyphens]{xurl}
\usepackage{hyperref}
\usepackage{booktabs}
\usepackage{amsfonts}
\usepackage{amsmath}
\usepackage{nicefrac}
\usepackage{microtype}
\usepackage{xcolor}
\usepackage[most]{tcolorbox}

\newtcolorbox{defbox}[1][]{
    colback=gray!5,
    colframe=gray!40,
    boxrule=0.5pt,
    arc=2pt,
    left=6pt,right=6pt,top=4pt,bottom=4pt,
    title={#1},
    fonttitle=\bfseries\small,
    coltitle=black,
    colbacktitle=gray!15
}
\usepackage{multirow}
\usepackage{array}
\usepackage{enumitem}
\usepackage{tikz}
\usepackage{graphicx}
\usetikzlibrary{positioning,arrows.meta,shapes.geometric,calc,patterns,decorations.pathreplacing}

\definecolor{cgold}{RGB}{46,125,50}       
\definecolor{cpoison}{RGB}{198,40,40}     
\definecolor{cabstain}{RGB}{96,125,139}   
\definecolor{cdrift}{RGB}{245,166,35}     

\title{A Failure-Mode Benchmark for Polymorphic Sybil Poisoning in RAG}

\author{%
  Donghyun Lee \\
  Department of Computer Engineering \\
  Dongguk University \\
  Seoul, Republic of Korea \\
  \texttt{donghyun0215@dgu.ac.kr} \\
  \And
  Juntae Kim\thanks{Corresponding author.} \\
  Department of Computer Engineering \\
  Dongguk University \\
  Seoul, Republic of Korea \\
  \texttt{jkim@dongguk.edu} \\
}

\begin{document}

\maketitle

\begin{abstract}
We release a benchmark and failure-mode-aware evaluation framework
for grounded QA under coordinated retrieval poisoning. The
framework partitions reader outputs into four mutually exclusive
categories (\emph{gold}, \emph{hijack}, \emph{abstention},
\emph{drift}), with instance-level paired clean-to-poison transition
matrices and a Forced Exposure protocol isolating reader-side
conflict resolution from retrieval variance. We introduce
\emph{polymorphic sybil poisoning}, a coordinated attack class in
which $S$ lexically diverse passages jointly support an
attacker-chosen target while evading lexical near-duplicate filters
that fully detect monomorphic baselines (capturing the residual
14.2\% with E5 cosine raises false-positive rate 9$\times$ on
legitimate same-topic pairs). A monomorphic--polymorphic ablation
under Forced Exposure isolates the diversity dimension and reveals
a $+$18.8pp hijack amplification (95\% paired bootstrap CI
$[+15.4, +22.4]$, $B{=}5{,}000$): monomorphic copies register only
4.0\% as hijack while polymorphic surface diversity recovers
22.8\%---a 5.7$\times$ amplification of the ASR-visible attack
channel. ASR alone treats every non-target output identically;
under attack, abstention and drift together hold 47--66\% of
output mass, unmonitored by ASR+ACC, and two readers at nearly
identical ASR (within 0.2pp) differ by 16.5pp on abstention and
17.2pp on drift---failure profiles invisible to ASR. We release the
frozen benchmark (3{,}145 questions, 2{,}982 retained sybil groups;
$S{=}6$ chosen to dominate top-10 retrieval slots,
\S\ref{sec:setup}), the official four-way evaluator,
paired-transition utilities, and the Forced Exposure harness across
five readers (7B--120B), two retrievers, and two cross-validation
datasets (TriviaQA, 2Wiki), under CC~BY-SA~4.0 (data) and MIT
(software); release information in \S\ref{sec:release}.
\end{abstract}

\section{Introduction}
\label{sec:intro}

Retrieval-augmented generation (RAG) systems ground their answers
in externally retrieved evidence, making the retrieval corpus a
direct attack surface~\citep{zou2025poisonedrag,
chaudhari2024phantom, xue2024badrag}. Existing multi-passage
attacks~\citep{zou2025poisonedrag} generate adversarial texts
without an explicit lexical-diversity constraint; the resulting
passages exhibit incidental similarity from shared
retrieval-condition components and remain susceptible to
near-duplicate corpus hygiene. We contrast the polymorphic regime
against a \emph{monomorphic baseline} (lexically near-identical,
mean Token Jaccard $\approx 1.00$) constructed here as the
worst-case lexical-similarity limit (\S\ref{sec:defense-evasion}).

\begin{figure}[t]
\centering
\includegraphics[width=\linewidth]{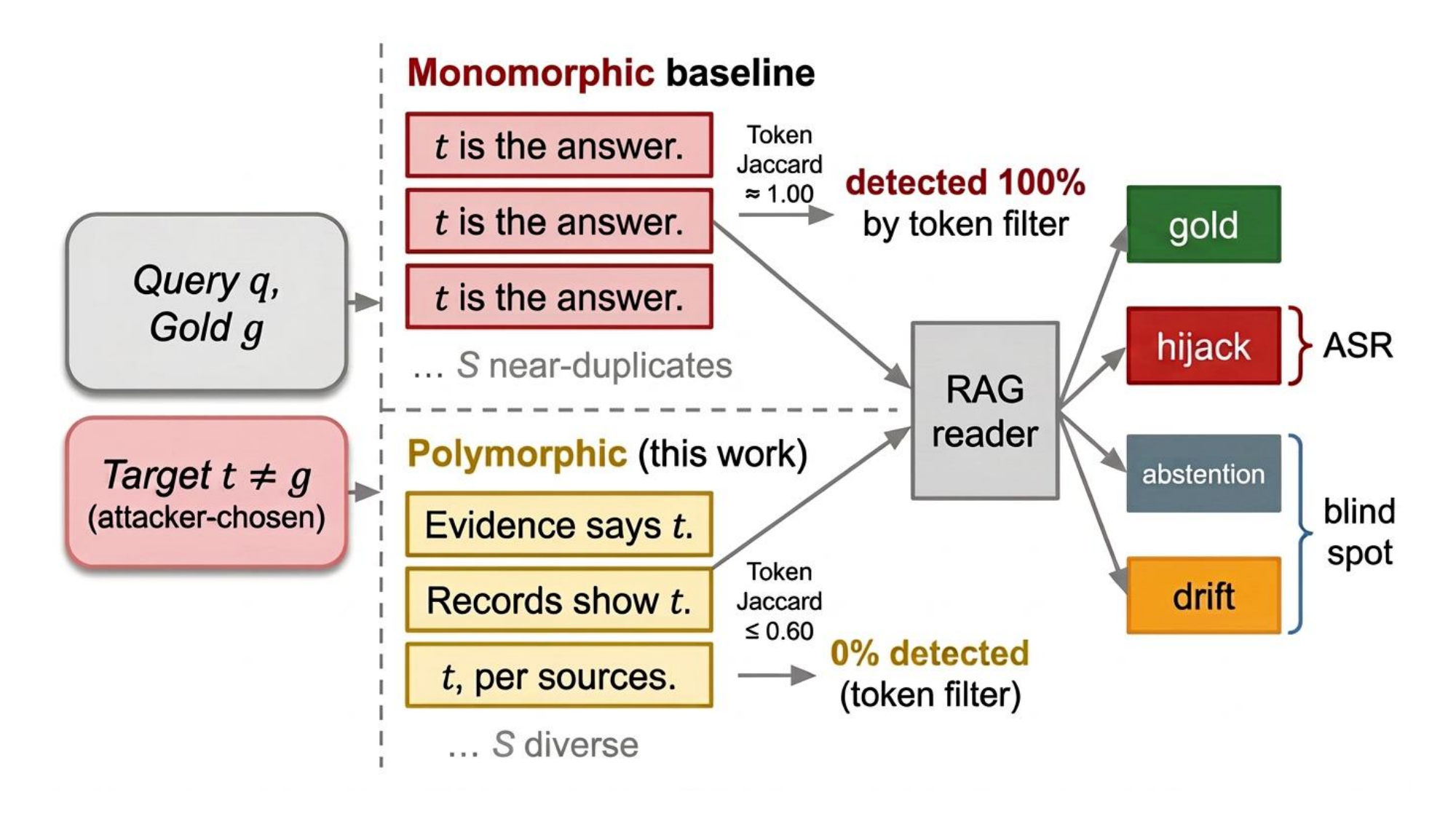}
\caption{\textbf{Polymorphic sybil poisoning vs.\ monomorphic
worst-case baseline.} Both inject $S{=}6$ passages supporting an
attacker-chosen target $t \neq g$. Sybil text shown is stylized;
full passages are $\sim$100-word natural-language narratives in
the released manifest (\S\ref{sec:construction},
\S\ref{app:meta}). Monomorphic (mean Token Jaccard $\approx 1.00$)
is constructed as the worst-case lexical-similarity limit and is
fully detected by a token-overlap filter at threshold $\geq 0.60$.
Polymorphic enforces $\tau_{\text{lex}}{=}0.8$ during generation
(achieved mean 0.32, max 0.60) and evades the same filter at 0\%
(embedding-filter trade-off: \S\ref{sec:defense-evasion}). Existing
multi-passage attacks~\citep{zou2025poisonedrag} fall in the
intermediate regime. Reader outputs redistribute across four
categories (gold/hijack/abstention/drift); abstention and drift
together account for 47--66\% under attack, invisible to
ASR$+$ACC.}
\label{fig:concept}
\end{figure}

We introduce \textbf{polymorphic sybil poisoning}
(Figure~\ref{fig:concept}): $S$ passages that jointly support an
attacker-chosen target answer, maintain low pairwise token overlap
($\tau_{\text{lex}}{=}0.8$ as a generation-time soft constraint;
achieved mean 0.32, max 0.60; \S\ref{sec:attack}), and pass a
verifier-LLM quality gate. The released benchmark fixes $S{=}6$,
chosen to dominate top-10 retrieval slots
(\S\ref{sec:setup}). A lexical near-duplicate
filter (Token Jaccard $\geq 0.60$) yields a binary separation
between polymorphic (0\%) and monomorphic (100\%) clusters;
embedding-based filtering admits no operating point matching this
gap (Table~\ref{tab:defense}, \S\ref{sec:defense-evasion}).
A monomorphic--polymorphic ablation under Forced Exposure isolates
the diversity dimension: monomorphic copies register only 4.0\% as
hijack while polymorphic surface diversity recovers 22.8\%---a
$+$18.8pp amplification (95\% paired bootstrap CI $[+15.4, +22.4]$,
$B{=}5{,}000$; \S\ref{sec:ablation}). Because polymorphic sybils
present diverse, seemingly independent evidence, reader outputs
under attack distribute across four categories---gold, hijack,
abstention, drift---rather than collapsing onto the hijack axis.
ASR captures only hijack; the remaining 47--66\% of output mass
falls outside both ASR and ACC. We propose a
\textbf{failure-mode-aware evaluation framework} with a four-way
partition, paired clean-to-poison transitions, and a Forced
Exposure protocol isolating reader-side conflict resolution.

\paragraph{Contributions.}
\textbf{(1)}~Polymorphic sybil poisoning, a coordinated attack
class defeating lexical near-duplicate filtering and forcing a
precision--recall trade-off on embedding-based filtering
(\S\ref{sec:attack}, \S\ref{sec:defense-evasion}).
\textbf{(2)}~A four-way evaluation framework with
\emph{instance-level paired clean-to-poison transition matrices}
and the \emph{Forced Exposure} protocol isolating reader-side
conflict resolution (\S\ref{sec:ablation}, \S\ref{sec:evaluation}).
\textbf{(3)}~A frozen benchmark (3{,}145Q, 2{,}982 retained;
cross-dataset on TriviaQA and 2WikiMultiHopQA) with an official
evaluator and reference results across five readers, two
retrievers, three conditions
(\S\ref{sec:results}, \S\ref{sec:analysis}).

\section{Related Work}
\label{sec:related}

\paragraph{Retrieval poisoning attacks.}
\citet{zhong2023poisoning} craft adversarial passages via
HotFlip-style perturbations to maximize retrieval similarity.
PoisonedRAG~\citep{zou2025poisonedrag} injects $N{=}5$ passages
per target question and reports 90\%+ ASR with Contriever+PaLM~2;
its black-box variant generates passages via LLM stochastic
sampling, yielding intermediate lexical diversity from sampling
randomness rather than explicit pairwise enforcement. Extensions
add trigger activation~\citep{chaudhari2024phantom},
backdoors~\citep{xue2024badrag}, and single-text
variants~\citep{zhang2025corruptrag}; all measure effectiveness
by ASR alone, leaving non-hijack failure modes unreported. Our
polymorphic class adds an \emph{explicit} pairwise-diversity
constraint ($\tau_{\text{lex}}$) and verifier gate, producing a
binary gap against lexical filters and shifting embedding-based
detection onto a precision--recall trade-off
(\S\ref{sec:defense-evasion}).

\paragraph{RAG evaluation frameworks.}
Prior frameworks evaluate RAG along complementary axes:
RGB~\citep{chen2024rgb}, RAGChecker~\citep{ru2024ragchecker},
SafeRAG~\citep{liang2025saferag},
PoisonArena~\citep{chen2025poisonarena}, and
ClashEval~\citep{wu2024clasheval}. RSB~\citep{zhang2025rsb} sweeps
13 attacks $\times$ 7 defenses reporting ACC, ASR, F1
independently---informationally equivalent to our four-way
partition in static reporting. Our novel component is the
\emph{instance-level paired transition matrix} resolving which
clean-correct outputs each attack converts into which failure
mode, combined with adversarial multi-document injection and
Forced Exposure.

\paragraph{Abstention in QA.}
AbstainQA~\citep{feng2024abstainqa} and
Abstain-QA~\citep{madhusudhan2025abstainqa} define benign-setting
abstention frameworks without adversarial injection. We decompose
their ``answered incorrectly'' category into targeted hijack
(attacker-intended) vs.\ drift (undirected)---a distinction
critical for defense design.

\section{Polymorphic Sybil Attack}
\label{sec:attack}

\subsection{Threat Model}

We adopt a gray-box threat model~\citep{zou2025poisonedrag}: the
attacker can \emph{inject} passages into the corpus but not modify
existing ones, has black-box access to reader and retriever, and
aims to make the system output an attacker-chosen target $t \neq g$
on question $q$.

\subsection{Formal Definition}

\begin{defbox}[Definition: Polymorphic Sybil Group]
Given a target question $q$ with gold answer $g$, a
\emph{polymorphic sybil group} is a tuple
$(q, g, t, P, S, \tau_{\text{lex}}, \theta_{\text{qc}})$ where
$t$ is the attacker-chosen target answer, $P = \{p_1, \ldots, p_S\}$
is a set of $S$ staged passages satisfying:
\begin{enumerate}[nosep,leftmargin=*]
\item \textbf{Target support}: each $p_i$ supports $t$ as the answer to $q$;
\item \textbf{Lexical diversity}: for all $i \neq j$,
$\text{TokenJaccard}(p_i, p_j) \leq \tau_{\text{lex}}$;
\item \textbf{Verifier gate}: the number of passages classified as
supporting \emph{only} the gold answer $g$ (and not $t$) is fewer
than $\theta_{\text{qc}}(S)$.
\end{enumerate}
\end{defbox}

In the released benchmark, $S{=}6$, $\tau_{\text{lex}}{=}0.8$, and
$\theta_{\text{qc}}(6){=}3$. $\tau_{\text{lex}}{=}0.8$ is a
generation-time soft constraint on the paraphrase-and-resample
loop (groups exceeding it are regenerated); the retained
distribution is much tighter (mean pairwise Token Jaccard 0.32,
max 0.60). Passages are generated by Llama-3.1-8B-Instruct
(temp$=$0.8) and verified by Qwen2.5-72B-Instruct.

\paragraph{Diversity axes.}
The definition operates on two axes: \emph{lexical} surface form
(Condition 2: $\tau_{\text{lex}}$ pairwise enforcement) and
\emph{group-level coordination} (Condition 3: verifier rejection
of groups whose constituent passages disagree on the target). Prior
multi-passage attacks~\citep{zou2025poisonedrag} produce surface
paraphrastic variation incidentally from sampling randomness but
constrain neither worst-case pairwise overlap nor within-group
target-support consistency. Semantic and evidence-source diversity
emerge as paraphrase-loop byproducts (\S\ref{sec:limitations}).

\subsection{Why Polymorphic Sybils Are Harder to Defend}
\label{sec:defense-evasion}

We compare polymorphic sybil groups ($\tau_{\text{lex}}{=}0.8$)
against the monomorphic worst-case baseline ($S{=}6$, verifier
gate, diversity filter disabled, 500-question subset; mean Token
Jaccard 1.00) under four corpus-level detection strategies
(Table~\ref{tab:defense}). Token Jaccard $\geq 0.60$ yields a
binary gap (mono 100\%, poly 0\%); E5 cosine flags monomorphic at
100\% at every threshold but trades polymorphic detection sharply
against FPR on legitimate same-topic pairs (capturing the residual
14.2\% at $\geq 0.95$ raises FPR from 0.2\% to 1.8\%, a 9$\times$
increase). Perplexity and query-paraphrase flag both variants at
$\geq 95$\% and do not separate them.

\begin{table}[t]
\caption{Defense detection rates: fraction of monomorphic vs.\
polymorphic sybil clusters flagged by each filter, with the
corresponding false-positive rate (FPR) on clean same-topic
passage pairs. Evaluated on the 500Q ablation subset (mean Token
Jaccard 0.27 on this subset; 0.32 on the full 2,982 release;
mono 1.00). Monomorphic clusters are detected at 100\% across all
embedding thresholds; polymorphic detection trades off sharply
against FPR.}
\label{tab:defense}
\centering
\small
\begin{tabular}{llrrr}
\toprule
Filter & Threshold & Mono det. & Poly det. & Same-topic FPR \\
\midrule
Token Jaccard      & $\geq 0.60$           & 100\% & 0\%   & --- \\
\midrule
E5 Cosine          & $\geq 0.90$           & 100\% & 98.6\% & 12.1\% \\
E5 Cosine          & $\geq 0.95$           & 100\% & 85.8\% & 1.8\% \\
E5 Cosine          & $\geq 0.99$           & 100\% & 2.0\%  & 0.2\% \\
\midrule
Perplexity (GPT-2) & $<100$                & 95.7\% & 99.0\% & --- \\
Query Paraphrase   & retain $\geq 1$ sybil & 100\%  & 100\%  & --- \\
\bottomrule
\end{tabular}
\end{table}

\subsection{Monomorphic vs.\ Polymorphic Outcome Redistribution}
\label{sec:ablation}

Beyond detection evasion, polymorphic diversity affects how the
reader \emph{responds} to the attack. Table~\ref{tab:ablation}
compares four-way outcome distributions under Forced Exposure on
the 500Q ablation subset (Qwen2.5-72B reader). The two retrievers
agree to three decimals (Forced Exposure pins placements
deterministically); generalization is discussed in
\S\ref{sec:limitations} (iv).

\begin{table}[t]
\caption{Monomorphic vs.\ polymorphic ($\tau_{\text{lex}}{=}0.8$)
outcome redistribution under Forced Exposure (500Q, Qwen2.5-72B;
both retrievers identical on point estimates to three decimals).
95\% paired bootstrap CIs ($B{=}5{,}000$, seed$=$42; ColBERT). All
four $\Delta$'s have CIs excluding zero---polymorphism produces a
coordinated four-channel redistribution rather than a
single-channel shift. Per-instance results in
\texttt{results/ablation\_paired\_ci.csv}.}
\label{tab:ablation}
\centering
\small
\begin{tabular}{lrrrr}
\toprule
Condition & Gold & Hijack & Abstention & Drift \\
\midrule
Monomorphic (no filter) & 0.164 & 0.040 & 0.382 & 0.414 \\
Polymorphic & 0.192 & 0.228 & 0.308 & 0.272 \\
\midrule
$\Delta$ (Poly $-$ Mono) & +0.028 & +0.188 & $-$0.074 & $-$0.142 \\
95\% CI & $[+0.006, +0.050]$ & $[+0.154, +0.224]$ & $[-0.104, -0.042]$ & $[-0.184, -0.100]$ \\
CI excludes 0 & yes & yes & yes & yes \\
\bottomrule
\end{tabular}
\end{table}

The two variants separate on all four channels under Forced
Exposure (all CIs excluding zero): polymorphic groups achieve
$+$18.8pp higher hijack, suppress abstention by $-$7.4pp, reduce
drift by $-$14.2pp, retain marginally more gold ($+$2.8pp). The
same six target positions delivered through monomorphic copies
register only 4.0\% as hijack despite 83.6\% of outputs deviating
from gold (mass absorbed by abstention 38.2\% and drift 41.4\%);
polymorphic surface diversity recovers 22.8\%---a 5.7$\times$
amplification of ASR-visible hijack.

\section{Benchmark Construction}
\label{sec:construction}

\subsection{Task, Pipeline, and Release Statistics}

The benchmark operates in a grounded QA setting: for each question
$q$, the system retrieves passages from a fixed corpus $C$ and
produces $\hat{y}$; evaluation uses strict exact match (EM) against
canonicalized
gold alias sets $G(q)$. The attacker selects a target answer
$t \notin G(q)$, constructs a polymorphic sybil group
(Definition~1), and injects all $S$ passages into $C$ without access
to reader parameters, retriever internals, or the evaluator.
Target answers are sampled from dataset-conditional plausible
alternatives; the target-sampling protocol is released with the code
package.

\paragraph{Source datasets and splits.}
Host questions are drawn from validation/dev splits of four
open-domain QA datasets (seed$=$42 fixed random samples):
NQ-open (\texttt{nq\_open}, 3{,}610$\to$1{,}145)~\citep{kwiatkowski2019naturalquestions,lee2019latent};
HotpotQA distractor (\texttt{hotpot\_qa/distractor}, 7{,}405$\to$2{,}000)~\citep{yang2018hotpotqa};
TriviaQA (\texttt{trivia\_qa/unfiltered.nocontext} val, 11{,}313$\to$2{,}000)~\citep{joshi2017triviaqa};
2WikiMultiHopQA (Alab-NII dev, 12{,}576$\to$3{,}000)~\citep{ho2020constructing}.
Exact identifiers and text are frozen in the released manifest with
SHA-256 checksums; reproducibility does not depend on upstream
snapshot stability.

The main manifest is built on NQ and HotpotQA via a three-stage
filter pipeline (all stage manifests released):
(1)~\emph{qc-v2} ($\tau_{\text{lex}}{=}0.6$ $+$ semantic verifier
$+$ target--gold disjoint check): 11{,}015 $\to$ 4{,}022;
(2)~\emph{class balance} (seed$=$42, HotpotQA capped at 2{,}000):
4{,}022 $\to$ 3{,}145 (1{,}145~NQ $+$ 2{,}000~HotpotQA);
(3)~\emph{strict release} ($\tau_{\text{lex}}{=}0.8$ $+$
$\theta_{\text{qc}}(6){=}3$): 3{,}145 $\to$ 2{,}982 (94.8\%;
96.6\%~NQ, 93.8\%~HotpotQA; mean pairwise Jaccard 0.32, max 0.60).
Generator: Llama-3.1-8B-Instruct ($Q8\_0$, temp$=$0.8); verifier:
Qwen2.5-72B-Instruct ($Q4\_K\_M$, temp$=$0)---distinct families to
avoid generator--verifier collapse; 5.2\% rejection at the strict
release filter indicates substantive filtering.

\paragraph{Cross-dataset validation layer.}
For generalizability we apply the same pipeline to
\textbf{TriviaQA} (2{,}000$\to$1{,}398; factoid; 69.9\% retention,
below 89.9--96.6\% elsewhere due to short factoid prompts giving
the verifier more headroom on ambiguous gold) and
\textbf{2WikiMultiHopQA} (3{,}000$\to$2{,}696; multi-hop). Released
as a separate layer (\S\ref{sec:cross-dataset}); a 2Wiki full-scale
18{,}000-passage QC audit is also released.

\section{Evaluation Framework}
\label{sec:evaluation}

\subsection{Operational Partition and Official Evaluator}
\label{sec:partition}

Let $\hat{y}$ denote the canonicalized evaluator output, $G$ the
canonicalized gold alias set, $T$ the canonicalized target alias
set, and $A$ the abstention marker set. We define four mutually
exclusive and collectively exhaustive categories:
\emph{gold retention} ($\hat{y} \in G$),
\emph{targeted hijack} ($\hat{y} \in T \setminus G$),
\emph{abstention} ($\hat{y} \in A$), and
\emph{third-answer drift} ($\hat{y} \notin G \cup T \cup A$).
This extends AbstainQA's four-scenario
framework~\citep{feng2024abstainqa} to the adversarial setting by
decomposing ``answered incorrectly'' into targeted hijack
(attacker-intended) and drift (undirected), a distinction critical
for defense evaluation.

The official evaluator applies strict EM against canonicalized
gold and target alias sets (lowercase, article strip,
non-alphanumeric removal) uniformly across tracks (full
canonicalization in App.~\ref{app:readers}). When an extracted
answer contains both gold and target aliases the evaluator raises
a \emph{conflict flag} resolved via first-entity rule; flag rates
are a descriptive signal varying by two orders of magnitude across
readers (per-reader breakdowns in output CSVs). A lenient
alias-substring variant raises ACC by 4.6--13.4pp but shifts
paired metrics by at most $\pm$2pp.

\subsection{Metrics, Paired Analysis, and Forced Exposure}
\label{sec:metrics}

\paragraph{Metrics and statistics.}
\textbf{ACC} (full-manifest gold), \textbf{ASR} (attack-condition
fraction in $T\setminus G$ on the retained-poison subset),
\textbf{abstention rate}, \textbf{drift rate}; paired metrics
(hijack, drift, attack-induced abstention) on the clean-correct,
retained-poison subset. We report 95\% percentile bootstrap CIs
($B{=}5{,}000$) on per-cell rates and paired bootstrap CIs on
differences (mono--poly $\Delta$, retriever $\Delta$); retriever
comparisons across 5 readers $\times$ 3 tracks (15 tests) use
Bonferroni $m{=}15$, $\alpha{=}0.0033$. Clean and attack/forced
outputs are matched on question identifier; conditional transition
matrices on the clean-correct subset quantify what fraction shifts
to each failure mode (\S\ref{sec:paired}).

\paragraph{Forced Exposure.}
The $S{=}6$ sybil passages are placed at protocol-defined top-10
positions alongside gold-supporting passages drawn under a shared
rule from the retained candidate pool. This eliminates retrieval
stochasticity and measures reader behavior when poison and gold
evidence coexist; empirical retrieval-reader decoupling is reported
in \S\ref{sec:retrieval-sanity}. Output schema contracts
(raw / eval / final / abstain / conflict-flag fields per question)
are specified in the released evaluator code.

\section{Experimental Setup}
\label{sec:systems}
\label{sec:setup}

The released benchmark fixes the sybil group size at $S{=}6$,
chosen to dominate top-10 retrieval slots under attack
while leaving room for organic gold-supporting passages elsewhere
in the candidate pool; $S$-sensitivity is left as future work.

Five main LLM readers span 7B--120B parameters across open and
proprietary families: Qwen2.5-7B/72B-Instruct~\citep{qwen2024qwen25},
Llama-3.1-70B-Instruct~\citep{dubey2024llama3}, GPT-OSS-120B
(Unsloth release; Harmony reasoning tags stripped post-hoc), and
GPT-4o-mini. Open-source readers run Q4\_K\_M GGUF via
\texttt{llama.cpp}; GPT-4o-mini uses the OpenAI Chat Completions
API. All readers share a byte-identical few-shot prompt (SHA-256
App.~\ref{app:readers}); decoding is temperature$=$0,
max\_tokens$=$128 except GPT-OSS-120B (max\_tokens$=$256 for
Harmony-format reasoning; truncation analysis
\S\ref{sec:cross-dataset}).

Retrieval uses Wikipedia DPR 100-word
split~\citep{karpukhin2020dpr} (21M passages). Main pipeline:
BM25$\to$ColBERTv2~\citep{santhanam2022colbertv2}
(top-1000$\to$top-10); cross-retriever:
E5-large-v2~\citep{wang2022e5} (FAISS) top-200 $\to$
\texttt{ms-marco-MiniLM-L6-v2} $\to$ top-10. Three conditions:
\textbf{clean} (organic); \textbf{attack} (frozen $P_q$ prepended
pre-rerank, displacing lowest-ranked organic); \textbf{Forced
Exposure} (top-10 = $S{=}6$ sybil $+$ 2 gold $+$ 2 organic filler;
slot-assignment rule in the harness). The 6:2:2 composition is
fixed a priori; ratio sensitivity is future work.

\section{Results}
\label{sec:results}

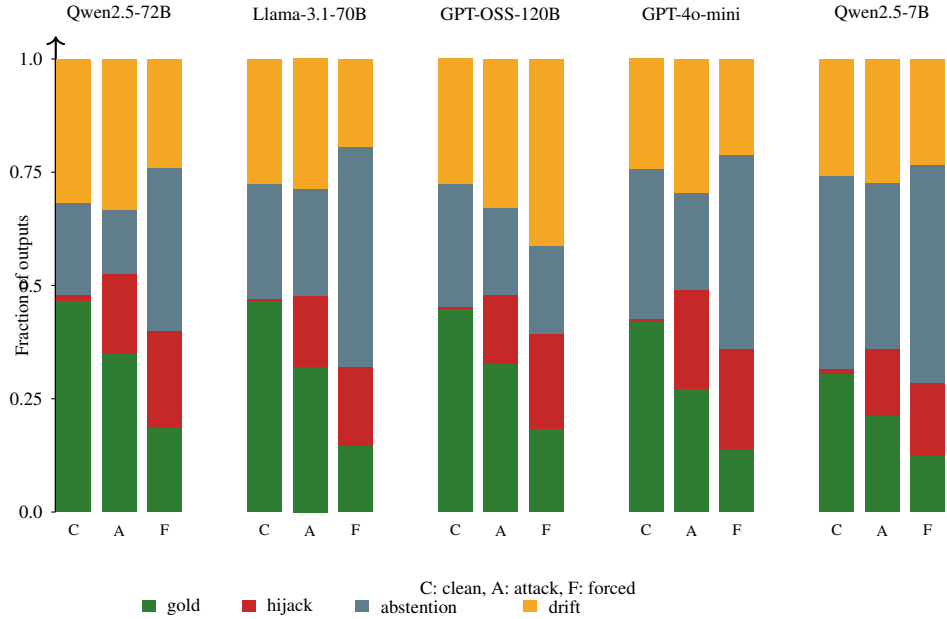
\begin{figure}[t]
\centering
\begin{tikzpicture}[x=6mm,y=60mm]
  \draw[->,thick] (-0.4,0) -- (-0.4,1.05) node[above,font=\footnotesize]{};
  \foreach \y/\lbl in {0.0/0.0, 0.25/0.25, 0.5/0.5, 0.75/0.75, 1.0/1.0}
    \draw (-0.5,\y) -- (-0.4,\y) node[left,font=\scriptsize,xshift=-0.05cm]{\lbl};
  \node[rotate=90,font=\scriptsize] at (-1.1,0.5) {Fraction of outputs};

  \foreach \idx/\rlabel/\xbase in {%
    0/Qwen2.5-72B/0,%
    1/Llama-3.1-70B/4.2,%
    2/GPT-OSS-120B/8.4,%
    3/GPT-4o-mini/12.6,%
    4/Qwen2.5-7B/16.8%
  }{
    \node[font=\scriptsize] at (\xbase+1,1.10) {\rlabel};
  }

  \foreach \x/\g/\p/\a/\d/\tlbl in {%
    0.0/0.467/0.012/0.204/0.317/C,%
    1.0/0.349/0.176/0.142/0.332/A,%
    2.0/0.189/0.212/0.360/0.238/F,%
    4.2/0.464/0.007/0.255/0.274/C,%
    5.2/0.320/0.158/0.235/0.287/A,%
    6.2/0.146/0.174/0.487/0.194/F,%
    8.4/0.449/0.005/0.270/0.276/C,%
    9.4/0.328/0.151/0.193/0.329/A,%
   10.4/0.184/0.210/0.195/0.410/F,%
   12.6/0.419/0.008/0.330/0.244/C,%
   13.6/0.273/0.217/0.215/0.295/A,%
   14.6/0.139/0.222/0.428/0.211/F,%
   16.8/0.306/0.010/0.427/0.257/C,%
   17.8/0.214/0.146/0.367/0.274/A,%
   18.8/0.123/0.161/0.483/0.233/F%
  }{
    \pgfmathsetmacro{\gtop}{\g}
    \pgfmathsetmacro{\ptop}{\g+\p}
    \pgfmathsetmacro{\atop}{\g+\p+\a}
    \fill[cgold]    (\x-0.38,0)      rectangle (\x+0.38,\gtop);
    \fill[cpoison]  (\x-0.38,\gtop)  rectangle (\x+0.38,\ptop);
    \fill[cabstain] (\x-0.38,\ptop)  rectangle (\x+0.38,\atop);
    \fill[cdrift]   (\x-0.38,\atop)  rectangle (\x+0.38,1.0);
    \node[font=\tiny] at (\x,-0.04) {\tlbl};
  }

  \begin{scope}[shift={(1.5,-0.22)}]
    \fill[cgold]    (0,0) rectangle (0.3,0.025); \node[anchor=west,font=\scriptsize] at (0.35,0.012) {gold};
    \fill[cpoison]  (2.2,0) rectangle (2.5,0.025); \node[anchor=west,font=\scriptsize] at (2.55,0.012) {hijack};
    \fill[cabstain] (4.7,0) rectangle (5.0,0.025); \node[anchor=west,font=\scriptsize] at (5.05,0.012) {abstention};
    \fill[cdrift]   (8.4,0) rectangle (8.7,0.025); \node[anchor=west,font=\scriptsize] at (8.75,0.012) {drift};
  \end{scope}
  \node[font=\scriptsize] at (10,-0.17) {C: clean, A: attack, F: forced};

\end{tikzpicture}
\caption{\textbf{Outcome redistribution under clean (C), attack
(A), and Forced Exposure (F)} across five readers (E5+CE,
Hotpot+NQ, $n{=}2{,}982$). Each bar sums to 1.0.
\textbf{ASR sees only the \textcolor{cpoison}{red (hijack)}
segment;} \textcolor{cabstain}{abstention} and
\textcolor{cdrift}{drift} together account for 47--66\% under
attack. GPT-OSS-120B vs.\ Qwen2.5-72B forced bars: similar hijack
heights (within 0.2pp) but inverted abstention/drift
(36.0/23.8 vs.\ 19.5/41.0). Llama-3.1-70B, GPT-4o-mini, and
Qwen2.5-7B shift mass primarily into abstention (42.8--48.7\%);
GPT-OSS-120B into drift (41.0\%), partly due to decode-budget
truncation (\S\ref{sec:cross-dataset}).}
\label{fig:redistribution}
\end{figure}

\begin{table}[t]
\caption{\textbf{Outcome redistribution under three conditions}
(Hotpot+NQ, $n{=}2{,}982$; rows sum to 1.000).
\textbf{Bold (E5+CE):} \emph{Clean Gold} ($\sim$31--47\%);
\emph{Forced Gold} ($\sim$12--19\%; drops $\sim$18--32pp from
clean); largest \emph{non-Gold forced} per reader marks where
displaced gold goes---abstention for 4 readers, drift for
GPT-OSS-120B---a difference invisible to ASR (forced hijack within
0.2pp for Qwen2.5-72B vs.\ GPT-OSS-120B yet abstention/drift
inverted: 0.360/0.238 vs.\ 0.195/0.410).
$^{\dagger}$GPT-4o-mini$\times$ColBERT attack on
$n_{\text{eff}}{=}2{,}711$ (9.1\% MCAR; App.~\ref{app:mcar});
paired analysis fixes all three conditions to this subset
($\leq 0.4$pp deviation between retrievers in forced cells).}
\label{tab:main-grid}
\centering
\small
\begin{tabular}{llrrrrr}
\toprule
Reader & Retriever & Track & Gold & Hijack & Abstention & Drift \\
\midrule
\multirow{6}{*}{Qwen2.5-72B}
  & \multirow{3}{*}{E5+CE}   & clean  & \textbf{0.467} & 0.012 & 0.204 & 0.317 \\
  &                          & attack & 0.349 & 0.176 & 0.142 & 0.332 \\
  &                          & forced & \textbf{0.189} & \textbf{0.212} & \textbf{0.360} & \textbf{0.238} \\
\cmidrule(l){2-7}
  & \multirow{3}{*}{ColBERT} & clean  & 0.435 & 0.014 & 0.232 & 0.319 \\
  &                          & attack & 0.330 & 0.177 & 0.161 & 0.332 \\
  &                          & forced & 0.189 & 0.212 & 0.360 & 0.238 \\
\midrule
\multirow{6}{*}{Llama-3.1-70B}
  & \multirow{3}{*}{E5+CE}   & clean  & \textbf{0.464} & 0.007 & 0.255 & 0.274 \\
  &                          & attack & 0.320 & 0.158 & 0.235 & 0.287 \\
  &                          & forced & \textbf{0.146} & 0.174 & \textbf{0.487} & 0.194 \\
\cmidrule(l){2-7}
  & \multirow{3}{*}{ColBERT} & clean  & 0.415 & 0.008 & 0.302 & 0.275 \\
  &                          & attack & 0.289 & 0.163 & 0.258 & 0.291 \\
  &                          & forced & 0.146 & 0.174 & 0.487 & 0.194 \\
\midrule
\multirow{6}{*}{GPT-OSS-120B}
  & \multirow{3}{*}{E5+CE}   & clean  & \textbf{0.449} & 0.005 & 0.270 & 0.276 \\
  &                          & attack & 0.328 & 0.151 & 0.193 & 0.329 \\
  &                          & forced & \textbf{0.184} & \textbf{0.210} & \textbf{0.195} & \textbf{0.410} \\
\cmidrule(l){2-7}
  & \multirow{3}{*}{ColBERT} & clean  & 0.410 & 0.007 & 0.297 & 0.286 \\
  &                          & attack & 0.296 & 0.158 & 0.221 & 0.325 \\
  &                          & forced & 0.184 & 0.210 & 0.195 & 0.410 \\
\midrule
\multirow{6}{*}{GPT-4o-mini}
  & \multirow{3}{*}{E5+CE}   & clean  & \textbf{0.419} & 0.008 & 0.330 & 0.244 \\
  &                          & attack & 0.273 & 0.217 & 0.215 & 0.295 \\
  &                          & forced & \textbf{0.139} & 0.222 & \textbf{0.428} & 0.211 \\
\cmidrule(l){2-7}
  & \multirow{3}{*}{ColBERT} & clean  & 0.379 & 0.008 & 0.378 & 0.236 \\
  &                          & attack$^{\dagger}$ & 0.224 & 0.192 & 0.307 & 0.276 \\
  &                          & forced & 0.143 & 0.222 & 0.427 & 0.208 \\
\midrule
\multirow{6}{*}{Qwen2.5-7B}
  & \multirow{3}{*}{E5+CE}   & clean  & \textbf{0.306} & 0.010 & 0.427 & 0.257 \\
  &                          & attack & 0.214 & 0.146 & 0.367 & 0.274 \\
  &                          & forced & \textbf{0.123} & 0.161 & \textbf{0.483} & 0.233 \\
\cmidrule(l){2-7}
  & \multirow{3}{*}{ColBERT} & clean  & 0.279 & 0.009 & 0.458 & 0.253 \\
  &                          & attack & 0.194 & 0.148 & 0.385 & 0.273 \\
  &                          & forced & 0.123 & 0.161 & 0.483 & 0.233 \\
\bottomrule
\end{tabular}
\end{table}

\paragraph{Outcomes redistribute beyond hijack under both attack
and Forced Exposure.}
Across five readers and both retrievers
(Table~\ref{tab:main-grid}, Fig.~\ref{fig:redistribution}),
clean-to-attack gold drops span 8.5--15.5pp; hijack gains $+$13.5
to $+$20.9pp; abstention shifts $-$1.9 to $-$11.5pp; drift shifts
$+$1.3 to $+$5.3pp. Abstention and drift together account for
47.4--65.9\% of outputs under attack---unmonitored by ACC$+$ASR.
Under \emph{Forced Exposure} the ASR-blindness sharpens: two
readers can land at near-identical hijack yet differ sharply on
which non-target channel absorbs the displaced mass (Qwen2.5-72B
vs.\ GPT-OSS-120B on Hotpot+NQ$\times$E5+CE: hijack 0.212/0.210
but abstention 0.360/0.195 and drift 0.238/0.410---a 16.5pp/17.2pp
inversion at near-identical hijack). The attack$\to$forced shift
moves mass primarily into abstention or drift in reader-specific
patterns; ASR-visible hijack remains tightly clustered.

\section{Analysis}
\label{sec:analysis}

\subsection{Paired Clean-to-Forced Transitions}
\label{sec:paired}

\begin{table}[t]
\caption{Clean-gold $\to$ Forced-Exposure transitions
($g{=}$gold, $p{=}$hijack, $a{=}$abstention, $d{=}$drift). For
instances classified as gold under clean, the fraction whose
forced output fell into each category, for all five main readers
and both retrievers.}
\label{tab:paired-forced}
\centering
\small
\begin{tabular}{llrrrr}
\toprule
Reader & Retr. & g$\to$g & g$\to$p & g$\to$a & g$\to$d \\
\midrule
\multirow{2}{*}{Qwen2.5-72B}
  & E5+CE   & 0.304 & 0.262 & 0.314 & 0.120 \\
  & ColBERT & 0.309 & 0.263 & 0.305 & 0.122 \\
\multirow{2}{*}{Llama-3.1-70B}
  & E5+CE   & 0.225 & 0.210 & 0.464 & 0.101 \\
  & ColBERT & 0.232 & 0.216 & 0.450 & 0.103 \\
\multirow{2}{*}{GPT-OSS-120B}
  & E5+CE   & 0.291 & 0.266 & 0.173 & 0.271 \\
  & ColBERT & 0.306 & 0.280 & 0.155 & 0.259 \\
\multirow{2}{*}{GPT-4o-mini}
  & E5+CE   & 0.229 & 0.269 & 0.390 & 0.111 \\
  & ColBERT & 0.239 & 0.286 & 0.355 & 0.120 \\
\multirow{2}{*}{Qwen2.5-7B}
  & E5+CE   & 0.217 & 0.220 & 0.410 & 0.153 \\
  & ColBERT & 0.219 & 0.221 & 0.398 & 0.162 \\
\bottomrule
\end{tabular}
\end{table}

Three reader profiles emerge (Table~\ref{tab:paired-forced},
Fig.~\ref{fig:redistribution}). \emph{Abstention-dominant}
(Qwen2.5-7B, Llama-3.1-70B, GPT-4o-mini, Qwen2.5-72B): forced
abstention 36.0--48.7\% on Hotpot+NQ over drift 19.4--23.8\%, with
the same ordering replicated on TriviaQA and 2Wiki.
\emph{Drift-dominant} (GPT-OSS-120B): drift 41.0\% on Hotpot+NQ
over abstention 19.5\%, drift 31.1\% on TriviaQA over abstention
12.0\% (consistent with 14--23\% Harmony-format truncation,
\S\ref{sec:cross-dataset}). \emph{Highest gold preservation}:
Qwen2.5-72B preserves the most gold on Hotpot+NQ and TriviaQA
(18.9\% and 29.0\%); 2Wiki gold is uniformly low (6.9--10.4\%).
Aggregate ACC drops obscure these splits: Qwen2.5-72B vs.\
GPT-OSS-120B differ by 0.2pp in forced hijack but 16.5pp in
abstention and 17.2pp in drift on Hotpot+NQ$\times$E5+CE---a
qualitative profile difference invisible to ASR$+$ACC.

\subsection{Retrieval Substrate Decoupling}
\label{sec:retrieval-sanity}

Forced Exposure pins sybil and gold placements deterministically,
so $|\Delta(\text{E5+CE} - \text{ColBERT})| = 0$ holds by
construction; we verify this as a sanity check
($|\Delta| \leq 0.004$ across error-free cells, most exactly
$0.000$). The substantive finding is on clean/attack: E5+CE yields
1.9--4.9pp higher gold rates than ColBERT (Bonferroni $m{=}15$,
$\alpha{=}0.0033$), reflecting the retrieval choice's effect on
which passages reach the reader before the placement protocol takes
over.

\subsection{Cross-Dataset Validation}
\label{sec:cross-dataset}

\begin{table}[t]
\caption{Cross-dataset forced-exposure outcomes (E5+CE; ColBERT
within three decimals on error-free cells). Hotpot+NQ ($n{=}2{,}982$),
TriviaQA ($n{=}1{,}398$, factoid), 2Wiki ($n{=}2{,}696$, multi-hop).}
\label{tab:cross-dataset}
\centering
\small
\begin{tabular}{lcccccc}
\toprule
 & \multicolumn{2}{c}{Hotpot+NQ} & \multicolumn{2}{c}{TriviaQA} & \multicolumn{2}{c}{2Wiki} \\
\cmidrule(lr){2-3} \cmidrule(lr){4-5} \cmidrule(lr){6-7}
Reader & Hijack & Drift & Hijack & Drift & Hijack & Drift \\
\midrule
Qwen2.5-72B   & 0.212 & 0.238 & 0.273 & 0.152 & 0.211 & 0.159 \\
Llama-3.1-70B & 0.174 & 0.194 & 0.180 & 0.117 & 0.152 & 0.084 \\
GPT-OSS-120B  & 0.210 & 0.410 & 0.313 & 0.311 & 0.189 & 0.297 \\
GPT-4o-mini   & 0.222 & 0.211 & 0.363 & 0.162 & 0.203 & 0.124 \\
Qwen2.5-7B    & 0.161 & 0.233 & 0.280 & 0.203 & 0.123 & 0.124 \\
\bottomrule
\end{tabular}
\end{table}

Three replicated findings (Table~\ref{tab:cross-dataset}):
(i)~forced hijack ranges across the three datasets span 2.8--16.0pp
per reader (Llama-3.1-70B 2.8pp, Qwen2.5-72B 6.2pp, GPT-OSS-120B
12.4pp, Qwen2.5-7B 15.8pp, GPT-4o-mini 16.0pp)---a spectrum from
near-invariance to substantial dataset sensitivity;
(ii)~TriviaQA produces the highest mean forced hijack (28.2\%;
Hotpot+NQ 19.6\%, 2Wiki 17.6\%), suggesting factoid surfaces leave
readers more susceptible than multi-hop chains;
(iii)~retriever decoupling replicates
($|\Delta(\text{E5+CE} - \text{ColBERT})| \leq 0.004$ across
error-free cells). Reader profiles also survive: GPT-OSS-120B
shows the highest drift on every dataset (29.7--41.0\%),
Llama-3.1-70B the highest abstention (48.7--68.5\%), Qwen2.5-72B
the highest gold on Hotpot+NQ and TriviaQA (18.9\% and 29.0\%;
2Wiki gold uniformly low, 6.9--10.4\%).

\paragraph{GPT-OSS-120B decode-budget truncation.}
GPT-OSS-120B's Harmony output exceeds the 256-token decode budget
on 14.3--23.3\% of Hotpot+NQ, 10.3--12.7\% of TriviaQA, and
14.1--19.5\% of 2Wiki queries; the evaluator scores these
truncated outputs as drift, contributing to GPT-OSS-120B's
elevated drift fraction. We treat this as a reader-specific
output behavior under a fixed decode budget; other readers at
max\_tokens$=$128 do not exhibit comparable truncation.

\subsection{Clean Drift Audit}
\label{sec:drift-audit}

To separate genuine reader reasoning errors from evaluator-side
extraction artifacts, an LLM-based classifier (Qwen2.5-72B-Instruct,
also used as verifier and a main reader; \S\ref{sec:limitations}~(v))
labels 1{,}356 clean-condition drift samples on the reconciled
manifest as GENUINE (reasoning error), EXTRACTION (gold present
but canonical extraction fails), or DATASET (gold annotation
ambiguity); automatic composition 89.6\%/3.2\%/7.0\%
(Table~\ref{tab:drift-audit}, App.~\ref{app:drift}). One author
re-labeled a 250-instance stratified subsample blind to the
classifier; human composition 84.4\%/15.2\%/0.4\% with raw
agreement 82.8\%, Cohen's $\kappa{=}0.262$ (fair, Landis--Koch).
Excluding GPT-OSS-120B's truncated reasoning outputs raises raw
agreement to 88.5\% but lowers $\kappa$ to 0.194 due to rising
expected-by-chance agreement (76.7\% to 85.7\%) under a more
class-imbalanced GENUINE-dominant distribution, not deteriorating
alignment (Table~\ref{tab:irr-breakdown}). The classifier
under-detects EXTRACTION (auto 3.2\% vs.\ human 15.2\%) and
over-attributes DATASET, particularly on GPT-OSS-120B
(Harmony-format truncation scored as dataset-ambiguity: auto
19--21\% DATASET on GPT-OSS subsample vs.\ human 0\%); the audit's
GENUINE fraction remains an upper bound. Cross-reader drift
comparisons in Table~\ref{tab:main-grid} use the four-way evaluator
and are unaffected by classifier bias.

\section{Limitations and Ethics}
\label{sec:limitations}

\paragraph{Limitations.}
\textbf{(i)}~Single attack class with two enforced diversity axes
(lexical pairwise overlap, group-level target-support consistency);
semantic and evidence-source diversity emerge as paraphrase-loop
byproducts; generator/verifier are both LLMs (partially controlled
by distinct model families).
\textbf{(ii)}~Forced Exposure measures reader-side conflict
resolution rather than deployed effectiveness; the fixed 6:2:2
(sybil:gold:filler) composition maximizes contention and ablation
channel-shift magnitudes (\S\ref{sec:ablation}) depend on this
ratio.
\textbf{(iii)}~NQ/HotpotQA contamination risk, partially mitigated
by cross-dataset validation.
\textbf{(iv)}~Ablation uses only Qwen2.5-72B on 500Q against a
worst-case monomorphic baseline; generalization across readers
and intermediate baselines is open.
\textbf{(v)}~Qwen2.5-72B serves three roles (verifier, reader,
drift-origin classifier); verifier--reader overlap is partially
controlled by three non-Qwen readers, classifier role is not
(human-classifier $\kappa{=}0.262$ fair, raw 82.8\%; raw 88.5\%,
$\kappa{=}0.194$ excl.\ GPT-OSS; non-Qwen replication deferred to
camera-ready).
\textbf{(vi)}~GPT-4o-mini$\times$ColBERT attack retains 9.1\% MCAR
API errors (App.~\ref{app:mcar}); paired analysis fixes all three
conditions to $n_{\text{eff}}{=}2{,}711$, inducing $\leq 0.4$pp
deviation from retriever equality in forced cells. GPT-OSS-120B
cells contain 3--6 \texttt{[ERROR]} outputs per cell from
Harmony-format parsing (evaluated as drift by the official rule).
\textbf{(vii)}~A small fraction of retained groups contain a
residual generator-preamble fragment (one of six positions): main
32/2{,}982 (1.07\%), TriviaQA 10/1{,}398 (0.72\%), 2Wiki 34/2{,}696
(1.26\%). Manifest-level Jaccard unaffected within $\pm 0.001$.
Detector $+$ flagged IDs in
\texttt{audit/detect\_preamble\_fragments.py}.
\textbf{(viii)}~Two manifest-quality processes for v1.0:
(a)~target metadata reconciliation reassigned target answers to
the sybil-dominant entity in $\sim$33\% of qc-v2 groups (3{,}181
across four manifests; per-instance labels released);
(b)~release-stage audit ($n{=}100$ random, seed$=$4202) finds 76\%
strict / 92\% majority / 4\% degenerate (\S\ref{app:meta}); strict
subset hijack averages $+$3.6pp above full-sample, matching the
polymorphic ablation $\Delta{=}+18.8$pp direction
(\S\ref{sec:ablation}). All paper statistics use the
verifier-accepted reconciled manifest; per-instance audit labels
released for downstream filtering.

\paragraph{Ethics.}
Released artifacts contain adversarial passages for robustness
evaluation; the attack class adds no capabilities beyond prior
work~\citep{zou2025poisonedrag, zhong2023poisoning}. We release
evaluation-time materials and prompts but not orchestration tooling;
the Datasheet (\S\ref{app:datasheet}) excludes production
red-teaming and attack-model training.

\paragraph{Release.}
\label{sec:release}
The complete benchmark will be publicly released upon publication.
\textbf{Benchmark proper} (code repository): frozen manifest
(3{,}145Q, 2{,}982 retained, $\sim$8.6\,MB), four-way evaluator,
paired-transition utilities, generator/verifier prompts with
SHA-256, cross-dataset and ablation layers, drift audit, Datasheet
(\S\ref{app:datasheet}), Croissant metadata
(\texttt{manifest/croissant.json}). \textbf{Prebuilt retrieval
indexes} (BM25 $\sim$11\,GB, E5 FAISS $\sim$81\,GB) over Wikipedia
DPR 100w---redistributed for reproducibility convenience. ColBERTv2
weights load from the official upstream checkpoint. Repository and
dataset URLs will be added in a future revision.

\clearpage
\bibliographystyle{plainnat}
\bibliography{references}

\appendix

\section{Artifact Specifications and Reproducibility}
\label{app:artifacts}

\subsection{Verifier Configuration}
\label{app:verifier}

Verifier $V$ is Qwen2.5-72B-Instruct GGUF (Q4\_K\_M;
\texttt{bartowski/Qwen2.5-72B-Instruct-GGUF}) served via
\texttt{llama.cpp} (context 8{,}192; temperature$=$0;
max\_tokens$=$32). The prompt (SHA-256 \texttt{0a4a02c9\ldots})
elicits a per-passage structured binary judgment
(\texttt{supports\_gold}, \texttt{supports\_target}). For a
candidate group $P{=}\{p_1,\ldots,p_S\}$, let
$n_{\text{gold-only}} = |\{i: \texttt{supports\_gold}_i \land
\lnot \texttt{supports\_target}_i\}|$; we retain $P$ iff
$n_{\text{gold-only}} < \theta_{\text{qc}}(S)$, with
$\theta_{\text{qc}}(6){=}3$ (implementation:
\texttt{verifier\_acceptance.py}).

\subsection{Polymorphic Sybil Generation Pipeline}
\label{app:generator}

Sybil passages are generated by Llama-3.1-8B-Instruct
(\texttt{Meta-Llama-3.1-8B-Instruct-Q8\_0.gguf}) via a
\texttt{llama.cpp} completion endpoint, deliberately distinct
from the Qwen2.5-72B verifier to avoid generator--verifier identity
collapse (the observed 5.2\% verifier rejection rate, well above
0\%, indicates substantive filtering rather than rubber-stamping).
For each manifest question $q$, the generator produces $S{=}6$
candidate passages supporting a target answer $t \notin G(q)$;
target answers are sampled from dataset-conditional plausible
alternatives (target-sampling code and seed released). The
generator prompt (SHA-256 \texttt{37eb61bc\ldots}) instructs the
model to produce lexically diverse passages supporting the target
while avoiding verbatim copies; an LLM-based paraphrase stage
(temperature$=$0.8, default top-p, held constant across questions
and datasets) yields lexically distinct variants from a shared
semantic template.

\subsection{Reader Configurations}
\label{app:readers}

All readers share a common few-shot QA template
(\texttt{COMMON\_QA\_SYSTEM\_PROMPT}) with three concise demonstrations
(1--3 token answers) plus an \texttt{Unknown} demonstration anchoring
abstention; byte-identical across readers, SHA-256
\texttt{bfda716d\ldots}.
Local GGUF readers use \texttt{llama.cpp} text completion;
GPT-OSS-120B uses \texttt{llama.cpp} chat completion with
Harmony-format post-processing; GPT-4o-mini uses the OpenAI Chat
Completions API. Each inference run emits per question the five
canonical fields \texttt{answer\_raw} / \texttt{answer\_eval} /
\texttt{answer\_final} / \texttt{abstain} / \texttt{conflict\_flag}
supporting deterministic re-scoring
(Table~\ref{tab:readers}).

\begin{table}[h]
\caption{Reader configurations. Decoding: temperature$=$0,
max\_tokens$=$128 (GPT-OSS-120B: max\_tokens$=$256). GGUF file
SHA-256 values in \S\ref{app:checksums}.}
\label{tab:readers}
\centering
\scriptsize
\setlength{\tabcolsep}{4pt}
\begin{tabular}{@{}llllr@{}}
\toprule
Reader & GGUF / Identifier & Backend & Quant.\ / Ctx. & License \\
\midrule
Qwen2.5-7B-Instruct
  & \texttt{bartowski/\allowbreak Qwen2.5-7B-Instruct-GGUF} & llama.cpp text & Q4\_K\_M, 8192 & Apache 2.0 \\
Qwen2.5-72B-Instruct
  & \texttt{bartowski/\allowbreak Qwen2.5-72B-Instruct-GGUF} & llama.cpp text & Q4\_K\_M, 8192 & Qwen License \\
Llama-3.1-70B-Instruct
  & \texttt{bartowski/\allowbreak Meta-Llama-3.1-70B-Instruct-GGUF} & llama.cpp text & Q4\_K\_M, 8192 & Llama 3.1 Comm. \\
GPT-OSS-120B
  & \texttt{unsloth/\allowbreak gpt-oss-120b-GGUF} & llama.cpp chat & Q4\_K\_M, 131K & Apache 2.0 \\
GPT-4o-mini
  & \texttt{gpt-4o-mini-2024-07-18}$^{\dagger}$ & OpenAI API & ---, 128K & OpenAI ToS \\
\bottomrule
\end{tabular}
\end{table}

$^{\dagger}$Observed snapshot that the \texttt{gpt-4o-mini} alias
resolved to during our runs; the alias itself is not guaranteed to
route to this snapshot for future callers (see
\S\ref{sec:limitations}).

\subsection{Pipeline Specification}
\label{app:pipeline}

\paragraph{Retrieval.}
Corpus: \texttt{wikipedia-dpr-100w}, 21{,}015{,}324 passages (DPR
conventions). \textbf{Main substrate:} Lucene BM25 ($\sim$11\,GB
index) top-1000 $\to$ ColBERTv2 (dim 128; compressed MaxSim index
from HF hub) rerank $\to$ top-10. \textbf{Cross-retriever substrate:}
E5 (\texttt{intfloat/e5-large-v2}, dim 1{,}024, FAISS
\texttt{IndexFlatIP}, $\sim$81\,GB) top-200, reranked by the
\texttt{ms-marco-MiniLM-L6-v2} cross-encoder ($\sim$22.7M params)
$\to$ top-10.

\paragraph{Sybil injection.}
Attack: frozen group $P_q$ prepended to the candidate pool before
reranking, displacing the lowest-ranked organic candidates to
maintain the retriever-specific candidate cap (top-1000 for BM25,
top-200 for E5). Forced Exposure: top-10 slots are deterministically
composed as 6 sybil passages, 2 gold-supporting passages drawn from
the retained candidate pool, and 2 organic filler passages from the
same pool; the slot-assignment rule (consistent with \S\ref{sec:setup})
is applied identically across retrievers and readers and is included
in the released evaluation harness.

\paragraph{Note on the \texttt{sybil\_\allowbreak in\_\allowbreak top10} diagnostic.}
The released artifacts include a per-query counter
\texttt{sybil\_\allowbreak in\_\allowbreak top10} that records, for
each query, the number of sybil passages present in the top-10
slots delivered to the reader. Under attack on Hotpot+NQ, the mean
of this counter is 4.000 for ColBERT (100\% of queries contain
$\geq 1$ sybil) and 3.75 for E5+CE (91.9\% of queries contain
$\geq 1$ sybil), confirming that both retrievers expose readers to
the injected sybils at comparable rates. Gold-supporting passages
reach the top-10 in 90.2\% of queries under both retrievers.
Under Forced Exposure the count is 6 by construction (sybil
placement is deterministic per the protocol; \S\ref{sec:metrics});
the released runner does not recompute the diagnostic in this
condition and emits \texttt{0.000} as a not-applicable placeholder
in the per-query log---this value indicates ``not measured,'' not
``not exposed.''

\subsection{Checksums and Licenses}
\label{app:checksums}

\paragraph{Checksums.}
Prompt-level SHA-256 is fixed: verifier \texttt{0a4a02c9\ldots},
common reader \texttt{bfda716d\ldots}, generator
\texttt{37eb61bc\ldots}. File-level checksums for all data and
software artifacts are committed as \texttt{SHA256SUMS} in the
release repository (verification: \texttt{sha256sum -c SHA256SUMS}).

\paragraph{Licenses.}
Main-release data (manifest, poison artifacts, monomorphic baseline)
under \textbf{CC~BY-SA~4.0} (share-alike inherited from NQ and
HotpotQA); software under \textbf{MIT}; TriviaQA samples and
2WikiMultiHopQA samples under \textbf{Apache~2.0}
(inherited from upstream). Users redistributing must comply with
the upstream source license in addition.

\subsection{Audit, Compute, and LLM Usage}
\label{app:meta}

\paragraph{Manual audit.}
Authors conducted blind audits at two pipeline stages.
\emph{Stage~1 (qc-v2):} 100 accepted and 50 rejected groups from
the qc-v2 pool ($n{=}4{,}022$, post semantic-verifier; seed$=$42)
scored under a framing-aware criterion (passage substantively
presents $t$ as the answer to $q$ in context); composition: 61\%
strict, 86\% majority, 12\% degenerate.
\emph{Stage~2 (release):} an independent 100-group audit drawn
directly from the released manifest
(\texttt{frozen\_manifest.jsonl}, $n{=}2{,}982$; seed$=$4202,
balanced 51 hotpot $+$ 49 nq) yields 76\% strict, 92\% majority,
4\% degenerate. The release-stage strict rate is higher because
class-balance and strict-$\tau_{\text{lex}}{=}0.8$ filtering steps
(\S\ref{sec:limitations}~(viii)) preferentially remove
lower-quality groups. On the strict-support subset of the release
audit ($n{=}76$), forced hijack averages $+$3.6pp above
full-sample across the five readers (Qwen2.5-72B $+$7.7,
Llama-3.1-70B $+$5.0, GPT-4o-mini $+$4.1, GPT-OSS-120B $+$2.7,
Qwen2.5-7B $-$1.7), with strong readers shifting from drift to
hijack ($-$4 to $-$6pp drift)---direction matches the polymorphic
ablation $\Delta{=}+18.8$pp (\S\ref{sec:ablation}). Per-instance
audit labels released
(\texttt{audit/sybil\_qc\_human\_audit.json},
\texttt{sybil\_qc\_v2\_release\_human.json},
\texttt{results/strict\_patch\_v2\_release.csv}).

\paragraph{Compute.}
2$\times$ NVIDIA A100 80GB GPUs for generation, verification, and
inference; CPU retrieval indexing with $\geq 256$\,GB RAM (the
reference machine had 376\,GiB physical RAM; FAISS index loading
and Pyserini BM25 dominate at $\sim$120\,GB peak resident, with the
remainder reserved for OS and runner overhead). Approximate
wall-clock: generation 48 GPU-h, verification 6 GPU-h, main-grid
inference 96 GPU-h, cross-dataset 30 GPU-h, ablation 4 GPU-h
(total $\sim$200 GPU-h).

\paragraph{LLM usage.}
LLMs serve as (1)~QC verifier (Qwen2.5-72B; \S\ref{app:verifier}),
(2)~sybil generator (Llama-3.1-8B; \S\ref{app:generator}),
(3)~main readers (\S\ref{app:readers}), and (4)~drift-origin audit
classifier (Qwen2.5-72B; \S\ref{app:drift}). Qwen2.5-72B's
verifier--reader--classifier triple role is discussed in
\S\ref{sec:limitations}. Any LLM use for writing or editing did
not affect the core methodology and is therefore not separately
declared, consistent with the NeurIPS policy.

\section{Drift Origin Audit Details}
\label{app:drift}

We sample 1{,}356 clean-condition drift instances on the reconciled
manifest (post-Phase-1 pool, seed$=$42); per-cell counts vary
(Table~\ref{tab:drift-audit}) reflecting per-cell drift incidence.
An LLM classifier (Qwen2.5-72B-Instruct \texttt{Q4\_K\_M},
temperature$=$0; same model as the construction verifier,
\S\ref{app:verifier}) labels each as \textbf{GENUINE}
(substantively different answer), \textbf{EXTRACTION} (gold present
but canonical extraction fails), or \textbf{DATASET} (gold
annotation incorrect). The use of Qwen2.5-72B in this third role
is acknowledged as a limitation (\S\ref{sec:limitations}, item
v).

\paragraph{Inter-rater agreement.}
To validate the automatic classifier, one author re-labeled a
250-instance subsample stratified across all 10 reader$\times$
retriever cells (seed$=$42), blind to the classifier output.
Human-labeled composition is 84.4\% GENUINE / 15.2\% EXTRACTION /
0.4\% DATASET, against automatic 89.6\% / 3.2\% / 7.0\% (post-Phase-1
reconciled manifest); raw agreement is 82.8\% and Cohen's
$\kappa{=}0.262$ (fair agreement on the Landis--Koch scale).
Per-class disagreement is concentrated on GPT-OSS-120B: when the 50
GPT-OSS-120B subsample instances (across both retrievers) are
excluded from the comparison, raw agreement rises to 88.5\% but
$\kappa$ falls to 0.194; this $\kappa$ decrease reflects rising
expected-by-chance agreement (76.7\% to 85.7\%) under a more
class-imbalanced GENUINE-dominant distribution, not a substantive
deterioration in classifier-human alignment. The classifier
under-detects EXTRACTION (auto 3.2\% vs.\ human 15.2\%, a 12.0pp
gap on the full sample) and over-attributes DATASET-issue
(auto 7.0\% vs.\ human 0.4\%, a 6.6pp gap), particularly on
GPT-OSS-120B where the classifier scores Harmony-format truncated
reasoning as dataset-ambiguity (auto 19--21\% DATASET on GPT-OSS
subsample vs.\ human 0\%). The audit's GENUINE fraction therefore
remains an upper bound; the human-labeled estimate is closer to
84\% GENUINE / 15\% EXTRACTION. Per-instance disagreement
breakdowns, per-class confusion matrices, and the labeling protocol
are released with the audit artifacts. This audit does not modify
the official evaluator and is independent of the four-way main-grid
metric.

\begin{table}[h]
\caption{Inter-rater agreement breakdown on the 250-instance
human-relabeled subsample (reconciled manifest;
\S\ref{sec:limitations}~(viii)). ``Excl.\ GPT-OSS-120B'' removes
the 50 GPT-OSS-120B subsample instances (across both retrievers).
The full-set $\kappa$ is depressed by Harmony-format truncation
(\S\ref{sec:cross-dataset}) being scored as dataset-ambiguity by
the classifier and as canonical-extraction failure by the human
annotator; the excluded subset's lower $\kappa$ reflects a more
class-imbalanced distribution that inflates expected-by-chance
agreement, not a deterioration in alignment.}
\label{tab:irr-breakdown}
\centering
\small
\begin{tabular}{lrr}
\toprule
Metric & Full set & Excl.\ GPT-OSS-120B \\
\midrule
$n$ (instances)              & 250    & 200    \\
\midrule
Classifier GENUINE\%         & 90.4\% & 93.5\% \\
Classifier EXTRACTION\%      & 2.4\%  & 2.0\%  \\
Classifier DATASET\%         & 6.8\%  & 4.0\%  \\
Human GENUINE\%              & 84.4\% & 91.5\% \\
Human EXTRACTION\%           & 15.2\% & 8.0\%  \\
Human DATASET\%              & 0.4\%  & 0.5\%  \\
\midrule
Raw agreement                & 82.8\% & 88.5\% \\
Cohen's $\kappa$             & 0.262  & 0.194  \\
Landis--Koch interpretation  & fair   & slight \\
\bottomrule
\end{tabular}
\end{table}

\begin{table}[h]
\caption{Per-reader drift-origin audit under clean conditions on the
\emph{reconciled manifest} (\S\ref{sec:limitations}~(viii)),
\emph{automatic classifier output} ($n{=}1{,}356$ total). Cohen's
$\kappa{=}0.262$ on a 250-instance human-relabeled subsample
(Table~\ref{tab:irr-breakdown}); excluded \texttt{[ERROR]} outputs
are scored as drift by the official rule and propagate from
GPT-OSS-120B's Harmony-format truncation
(\S\ref{sec:cross-dataset}). Two \texttt{parse\_fail} instances on
Qwen2.5-72B (one per retriever) are excluded from the breakdown.}
\label{tab:drift-audit}
\centering
\small
\begin{tabular}{llrrrrr}
\toprule
Reader & Retr. & $n$ & GENUINE & EXTRACT & DATASET & GENUINE\% \\
\midrule
\multirow{2}{*}{Qwen2.5-7B}
  & E5+CE   & 117 &  98 &  8 & 11 & 83.8\% \\
  & ColBERT &  98 &  74 & 11 & 13 & 75.5\% \\
\multirow{2}{*}{Qwen2.5-72B}
  & E5+CE   & 158 & 154 &  2 &  1 & 97.5\% \\
  & ColBERT & 132 & 130 &  0 &  1 & 98.5\% \\
\multirow{2}{*}{Llama-3.1-70B}
  & E5+CE   & 139 & 135 &  2 &  2 & 97.1\% \\
  & ColBERT & 121 & 117 &  4 &  0 & 96.7\% \\
\multirow{2}{*}{GPT-4o-mini}
  & E5+CE   & 140 & 136 &  4 &  0 & 97.1\% \\
  & ColBERT & 116 & 113 &  3 &  0 & 97.4\% \\
\multirow{2}{*}{GPT-OSS-120B}
  & E5+CE   & 191 & 149 &  5 & 37 & 78.0\% \\
  & ColBERT & 144 & 109 &  5 & 30 & 75.7\% \\
\midrule
\textit{Total} & & 1{,}356 & 1{,}215 & 44 & 95 & 89.6\% \\
\bottomrule
\end{tabular}
\end{table}

\section{API-Error Missingness and MCAR Verification}
\label{app:mcar}

One main-grid cell (GPT-4o-mini$\times$ColBERT attack) retains
residual OpenAI API errors after rerun attempts (271 errors,
9.1\%). The forced-condition runs themselves were API-error-free in
the top-10 forced exposure pass; however, for paired
clean--attack--forced analysis on GPT-4o-mini$\times$ColBERT all
three cells are reported on the same $n_{\text{eff}}{=}2{,}711$
question subset (excluding the attack cell's error-affected
questions across all three conditions). This induces a
$\leq 0.4$pp deviation from the by-construction retriever equality
in GPT-4o-mini's forced cells. For the attack cell itself, validity
of rate estimates depends on missingness being uncorrelated with
reader outcome (MCAR).

\begin{table}[h]
\caption{API-error inventory for GPT-4o-mini$\times$ColBERT$\times$
attack. Clean-condition gold rate is computed on the same question
identifiers under the clean track (no errors). Pearson $\chi^{2}$
tests independence of clean-condition gold counts between
error-affected and error-free subsets.}
\label{tab:mcar}
\centering
\small
\begin{tabular}{lrrrrr}
\toprule
Cell & $n_{\text{err}}$ & $n_{\text{eff}}$ & Gold$_{\text{err}}$ & Gold$_{\text{non-err}}$ & $\chi^{2}$ $p$ \\
\midrule
ColBERT attack & 271 & 2{,}711 & 0.384 & 0.395 & 0.78 \\
\bottomrule
\end{tabular}
\end{table}

Clean-condition gold rates differ by 1.1pp between error and
non-error subsets (Table~\ref{tab:mcar}); $\chi^{2}$ fails to
reject independence ($p{=}0.78$), supporting MCAR at the resolution
of available evidence. MCAR is testable only on clean-condition
outcomes and is not directly verifiable on attack outcomes; the
verification is therefore necessary but not sufficient
(\S\ref{sec:limitations}, item vi). Full re-evaluation against a
more stable endpoint is deferred to the camera-ready revision.

\section{Bootstrap Confidence Intervals}
\label{app:ci}

95\% bootstrap CIs (percentile, $B{=}5{,}000$) for main-grid
four-way partition metrics are reported in
Table~\ref{tab:main-grid-ci}. Cross-dataset CI half-widths are
1.7--2.6pp on TriviaQA ($n{=}1{,}398$) and 1.0--1.9pp on 2Wiki
($n{=}2{,}696$); cross-reader differences discussed in
\S\ref{sec:cross-dataset} exceed CI overlap.

\begin{table}[h]
\caption{Main-grid 95\% bootstrap CIs (Hotpot+NQ, $n{=}2{,}982$).
Format: point estimate [CI lower, CI upper]. GPT-4o-mini$\times$ColBERT
attack uses an effective denominator
($n_{\text{eff}}{=}2{,}711$); see App.~\ref{app:mcar}.}
\label{tab:main-grid-ci}
\centering
\scriptsize
\begin{tabular}{lllllll}
\toprule
Reader & Retriever & Track & Gold [95\% CI] & Hijack [95\% CI] & Abstention [95\% CI] & Drift [95\% CI] \\
\midrule
Qwen2.5-72B & E5+CE & clean & 0.467 [0.449,0.484] & 0.012 [0.009,0.016] & 0.204 [0.189,0.218] & 0.317 [0.300,0.335] \\
Qwen2.5-72B & E5+CE & attack & 0.349 [0.333,0.366] & 0.176 [0.162,0.190] & 0.142 [0.129,0.155] & 0.332 [0.316,0.349] \\
Qwen2.5-72B & E5+CE & forced & 0.189 [0.175,0.204] & 0.212 [0.198,0.227] & 0.360 [0.343,0.378] & 0.238 [0.223,0.254] \\
Qwen2.5-72B & ColBERT & clean & 0.435 [0.418,0.453] & 0.014 [0.010,0.018] & 0.232 [0.217,0.247] & 0.319 [0.302,0.336] \\
Qwen2.5-72B & ColBERT & attack & 0.330 [0.313,0.347] & 0.177 [0.164,0.191] & 0.161 [0.147,0.174] & 0.332 [0.316,0.349] \\
Qwen2.5-72B & ColBERT & forced & 0.189 [0.175,0.203] & 0.212 [0.197,0.227] & 0.360 [0.343,0.379] & 0.238 [0.223,0.254] \\
\midrule
Llama-3.1-70B & E5+CE & clean & 0.464 [0.446,0.482] & 0.007 [0.004,0.011] & 0.255 [0.239,0.271] & 0.274 [0.258,0.290] \\
Llama-3.1-70B & E5+CE & attack & 0.320 [0.303,0.336] & 0.158 [0.145,0.171] & 0.235 [0.220,0.250] & 0.287 [0.271,0.303] \\
Llama-3.1-70B & E5+CE & forced & 0.146 [0.133,0.159] & 0.174 [0.161,0.187] & 0.487 [0.469,0.504] & 0.194 [0.180,0.208] \\
Llama-3.1-70B & ColBERT & clean & 0.415 [0.398,0.433] & 0.008 [0.005,0.011] & 0.302 [0.286,0.318] & 0.275 [0.259,0.291] \\
Llama-3.1-70B & ColBERT & attack & 0.289 [0.273,0.305] & 0.163 [0.150,0.176] & 0.258 [0.242,0.274] & 0.291 [0.275,0.307] \\
Llama-3.1-70B & ColBERT & forced & 0.146 [0.133,0.159] & 0.174 [0.160,0.188] & 0.487 [0.469,0.504] & 0.194 [0.180,0.208] \\
\midrule
GPT-OSS-120B & E5+CE & clean & 0.449 [0.431,0.467] & 0.005 [0.003,0.008] & 0.270 [0.254,0.286] & 0.276 [0.259,0.292] \\
GPT-OSS-120B & E5+CE & attack & 0.328 [0.311,0.345] & 0.151 [0.138,0.164] & 0.193 [0.179,0.207] & 0.329 [0.312,0.346] \\
GPT-OSS-120B & E5+CE & forced & 0.184 [0.170,0.199] & 0.210 [0.196,0.225] & 0.195 [0.181,0.210] & 0.410 [0.392,0.428] \\
GPT-OSS-120B & ColBERT & clean & 0.410 [0.392,0.428] & 0.007 [0.004,0.010] & 0.297 [0.281,0.315] & 0.286 [0.270,0.302] \\
GPT-OSS-120B & ColBERT & attack & 0.296 [0.280,0.313] & 0.158 [0.145,0.171] & 0.221 [0.206,0.236] & 0.325 [0.308,0.341] \\
GPT-OSS-120B & ColBERT & forced & 0.184 [0.171,0.199] & 0.210 [0.196,0.225] & 0.195 [0.181,0.209] & 0.410 [0.393,0.428] \\
\midrule
GPT-4o-mini & E5+CE & clean & 0.419 [0.401,0.436] & 0.008 [0.005,0.011] & 0.330 [0.313,0.347] & 0.244 [0.228,0.260] \\
GPT-4o-mini & E5+CE & attack & 0.273 [0.258,0.290] & 0.217 [0.202,0.231] & 0.215 [0.200,0.229] & 0.295 [0.278,0.311] \\
GPT-4o-mini & E5+CE & forced & 0.139 [0.127,0.152] & 0.222 [0.208,0.237] & 0.428 [0.410,0.445] & 0.211 [0.196,0.226] \\
GPT-4o-mini & ColBERT & clean & 0.379 [0.361,0.397] & 0.008 [0.005,0.011] & 0.378 [0.360,0.395] & 0.236 [0.220,0.251] \\
GPT-4o-mini & ColBERT & attack$^{\dagger}$ & 0.224 [0.209,0.239] & 0.192 [0.178,0.207] & 0.307 [0.291,0.324] & 0.276 [0.260,0.292] \\
GPT-4o-mini & ColBERT & forced & 0.143 [0.130,0.155] & 0.222 [0.208,0.237] & 0.427 [0.409,0.445] & 0.208 [0.194,0.223] \\
\midrule
Qwen2.5-7B & E5+CE & clean & 0.306 [0.290,0.323] & 0.010 [0.007,0.014] & 0.427 [0.409,0.444] & 0.257 [0.241,0.273] \\
Qwen2.5-7B & E5+CE & attack & 0.214 [0.199,0.230] & 0.146 [0.133,0.159] & 0.367 [0.349,0.384] & 0.274 [0.258,0.289] \\
Qwen2.5-7B & E5+CE & forced & 0.123 [0.112,0.135] & 0.161 [0.148,0.174] & 0.483 [0.465,0.500] & 0.233 [0.217,0.248] \\
Qwen2.5-7B & ColBERT & clean & 0.279 [0.263,0.295] & 0.009 [0.006,0.013] & 0.458 [0.441,0.476] & 0.253 [0.238,0.269] \\
Qwen2.5-7B & ColBERT & attack & 0.194 [0.180,0.208] & 0.148 [0.135,0.161] & 0.385 [0.368,0.403] & 0.273 [0.258,0.289] \\
Qwen2.5-7B & ColBERT & forced & 0.123 [0.112,0.136] & 0.161 [0.148,0.174] & 0.483 [0.465,0.501] & 0.233 [0.218,0.248] \\
\bottomrule
\end{tabular}
\end{table}

\section{Datasheet for Datasets}
\label{app:datasheet}

We follow the Datasheet for Datasets template (abridged; full
template in the release repository).

\paragraph{Motivation \& composition.}
The benchmark supports failure-mode-aware evaluation of RAG systems
under polymorphic sybil retrieval poisoning. Contents: 3{,}145
questions with paired polymorphic sybil groups ($S{=}6$; 2{,}982
retained). Main sources: NQ-open validation (1{,}145 sampled),
HotpotQA distractor dev (2{,}000 sampled). Validation layer:
TriviaQA \texttt{unfiltered.nocontext} validation (2{,}000 sampled;
1{,}398 retained), 2WikiMultiHopQA dev (3{,}000 sampled;
2{,}696 retained). All samples use seed$=$42 from the respective
pool (\S\ref{sec:construction}). Ablation: monomorphic baseline
(500Q subset). Target answers are drawn from dataset-conditional
plausible alternatives; sybils are generated by
Llama-3.1-8B-Instruct under the acceptance-filter protocol
(\S\ref{app:verifier}); aliases follow source conventions.

\paragraph{Uses, distribution, maintenance.}
Intended for evaluating RAG robustness to coordinated retrieval
poisoning; not intended for training attack or defense models (risk
of distribution collapse) or for red-teaming production systems.
Released via public repository under licenses in
\S\ref{app:checksums}; integrity verification via SHA-256.
Versioned frozen release; errata published as patch versions
without modifying the evaluator or official scoring rule.

\end{document}